\documentclass[
reprint,
superscriptaddress,
 amsfonts,
 amsmath,amssymb,
 aps,
 prd,
]{revtex4-2}
\usepackage{lineno}
\usepackage{subfig}
\usepackage{hyperref}
\usepackage{color}

\usepackage{graphicx}% Include figure files
\usepackage{bm}% bold math
\begin{document}

%\preprint{APS/123-QED}

\title{Primordial Black Holes Formed during Magneto-Hydrodynamic Turbulence in The Early Universe as Dominant Part of Dark Matter}

\author{Jia-Xiang Liang}
\email{Corresponding author: 220220939631@lzu.edu.cn}
\affiliation{International Centre for Theoretical Physics Asia-Pacific, University of Chinese Academy of Sciences, Beijing 100190, P. R. China}
\affiliation{Center for Gravitational Wave Experiment, National Microgravity Laboratory, Institute of Mechanics, Chinese Academy of Sciences, Beijing 100190, China}
\affiliation{Lanzhou Center of Theoretical Physics, Lanzhou University, Lanzhou 730000, China}

\author{Peng Xu}
\email{Corresponding author: xupeng@imech.ac.cn}
\affiliation{Center for Gravitational Wave Experiment, National Microgravity Laboratory, Institute of Mechanics, Chinese Academy of Sciences, Beijing 100190, China}
\affiliation{Lanzhou Center of Theoretical Physics, Lanzhou University, Lanzhou 730000, China}
\affiliation{Hangzhou Institute for Advanced Study, University of Chinese Academy of Sciences, Hangzhou 310024, China}
\affiliation{Taiji Laboratory for Gravitational Wave Universe (Beijing/Hangzhou), University of Chinese Academy of Sciences, Beijing 100049, China}

\author{Ming-hui Du}
\affiliation{Center for Gravitational Wave Experiment, National Microgravity Laboratory, Institute of Mechanics, Chinese Academy of Sciences, Beijing 100190, China}

\author{Zi-Ren Luo}
\affiliation{Center for Gravitational Wave Experiment, National Microgravity Laboratory, Institute of Mechanics, Chinese Academy of Sciences, Beijing 100190, China}
\affiliation{Hangzhou Institute for Advanced Study, University of Chinese Academy of Sciences, Hangzhou 310024, China}
\affiliation{Taiji Laboratory for Gravitational Wave Universe (Beijing/Hangzhou), University of Chinese Academy of Sciences, Beijing 100049, China}

\date{\today}% It is always \today, today,

\begin{abstract}
Primordial black holes (PBHs) offer a compelling candidate for dark matter. The production of PBHs through well-tested and accepted physical processes is highly worthy of investigation. This work highlights the role of turbulences in the very early universe in sustaining intense and persistent fluctuations in energy or mass density, which could provide a natural mechanism for PBH formation in the primordial universe. We analyze the mass range and abundance of PBHs produced in the magnetohydrodynamic turbulence induced by the electroweak phase transition. Remarkably, we find that the mass range of the produced PBHs falls within the most viable ``asteroid mass'' window from the present-day observations, and within natural parameter regions their abundance can be sufficiently large. These findings suggest that PBHs produced during magnetohydrodynamic turbulence in the very early universe may comprise a dominant part of dark matter.

% \begin{description}
% Secondary publications and information retrieval purposes.
% You may use the \texttt{description} environment to structure your abstract;
% use the optional argument of the \verb+\item+ command to give the category of each item. 
% \end{description}
\end{abstract}

\maketitle

\textit{Introduction}.--Numerous astrophysical and cosmological observations have provided us the irrefutable evidence that about 85\% of the matter in our universe is in the form of cold and nonbaryonic dark matter (DM) \cite{sahni20045,bertone2018history,efstathiou1990cosmological,clowe2006direct,khlopov2010primordial,2014}. The physical origin and nature of DM remain still one of the most perplexing mysteries in modern cosmology and physics, that has inspired a great deal of research. Among the DM candidates, primordial black hole (PBH) can be taken as the most appealing one that satisfies almost all the necessary requirements.

PBHs were proposed by Zel’dovich and Novikov \cite{zel1967hot}, and Hawking \cite{hawking1971gravitationally,carr1974black} as hyperthotical relics from the collapse of overdense regions in the very early universe. Supposed to form in the radiation-dominated era before nucleosynthesis, PBHs should naturally be considered as nonbaryonic, and today it can be viewed as the only surviving DM candidate that does not invoke new particles beyond the Standard Model. Though lacking direct and decisive evidence for their existence, PBHs nevertheless have drawn extensive attention in the past fifty years and become one of the most active research fields in astrophysics and cosmology, since their existence could have important implications and provide natural explanations for various observational conundra and challenges, including, most importantly, the DM puzzle. For example \cite{carr2024observational,escriva2024primordial}, it is suggested that the coalescing black hole binaries observed by the LIGO-Virgo-KAGRA collaborations might contain primordial ones and the unexpected mass gaps and low spins in these stellar mass black holes might be explained \cite{bird2016did,sasaki2016primordial,clesse2017clustering,belotsky2019clusters}, and also very large PBHs could provide seeds for supermassive black holes resided in galactic nuclei with high redshifts \cite{dolgov2023tension}. 

Since the encouraging results from the MACHO (Massive Compact Halo Object) project in the 1990s that DM might be compact objects with mass $\sim 0.5 M_{\odot}$ \cite{alcock1997macho}, severe constraints from joint observations and theoretical considerations \cite{carr2016primordial,lehoucq2009new,ade2016planck} had been imposed on the allowed mass range of PBHs. Today, attention has been shifted to the most viable ``asteroid mass'' window of $10^{17}\ g\sim 10^{24}\ g$, where the PBHs could comprise the dominant part of DM \cite{carr2016primordial,green2024primordial}. However, producing PBHs within a given mass range generally depends on primordial density fluctuations with enhancement peaked at the corresponding small scales, which could be achieved by the fine-tuned background dynamics of primordial quantum fields in the very early universe. Furthermore, to produce enough number of PBHs, such enhanced primordial fluctuations on certain scales have to be many orders of magnitude larger than those on cosmological scales, which can be achieved in certain inflation models. This introduces the ``beyond the Standard Model'' physics for PBH DM proposals, while at the same time brings also new challenges. Various alternative mechanisms for PBH formation have been proposed, including sound-speed resonance during inflation \cite{cai2018primordial}, scalar field instabilities \cite{khlopov1985gravitational}, collisions of highly energetic particles \cite{saini2018modified}, topological defects \cite{hawking1989black,polnarev1991formation}, and modified gravity scenarios \cite{barrow1996formation,kawai2021primordial}. Notably, cosmological phase transitions in the early Universe provide a rich environment for PBH production through multiple distinct channels. These include the gravitational collapse of false vacuum bubbles nucleated at a critical rate during first-order phase transitions \cite{hawking1982bubble}, PBH formation via delayed vacuum decay—where asynchronized phase transitions give rise to overdense regions \cite{liu2022primordial}—and the generation of large curvature perturbations through bubble collisions, which may exceed the threshold for gravitational collapse \cite{jung2024primordial}.

In this work, we emphasize the important role of turbulent behavior of the primordial plasma in the  very early universe, particularly the potential implications for PBH formation. 
The fluid model can be viewed as the most successful model for macroscopic and continuous matter systems. In nature, situations beyond the ideal fluid model, involving complicated disturbances, nonlinearities, dissipation, and instabilities, are more common, and turbulence is a rather prevalent state for fluids of high Reynolds number \cite{pope2001turbulent}. 
On the other hand, it had been noticed that the Reynolds number for the high-temperature and extremely (or even fully) relativistic fluids in the very early universe was expected to be very high \cite{giovannini2012reynolds,caprini2009stochastic, arnold2000transport}, and turbulences are believed to arise not only in the primordial fluids \cite{caprini2009stochastic,neronov2021nanograv}, but also even in the violent dynamical spacetime itself \cite{galtier2017turbulence}. 
Here, we focus on the magnetohydrodynamic (MHD) turbulence that persists for a period of many Hubble times ($H^{-1}$) during and after the first-order electroweak phase transition in the early universe \cite{caprini2009stochastic,peter2018gravitational,subramanian2016origin}.
The most noteworthy feature of such turbulences here is the enhanced and persistent fluctuations of energy or mass density, and as the density fluctuations of certain scales exceed a given threshold, PBHs within a certain mass range could then be formed. 
This therefore provides us a natural mechanism, that based on such well-accepted physical processes, for PBH formation in the very early universe.

In the following, we explain the basic principles of PBH formation from MHD turbulence induced by the electroweak phase transition and give estimations of the PBH mass range and abundance. 
%Meanwhile, we have calculated the abundance of PBHs for different parameter ranges and provided the constraints on these parameters.
Remarkably, it turns out that the mass range falls precisely within the most viable open asteroid mass window, and, within natural parameter regions of the MHD that match remarkably those from the previous and independent studies, the PBHs generated can comprise the dominant fraction of DM. 
%This strongly indicates a novel formation mechanism for the PBH DM, in which turbulence in primordial fluids plays a critical role.  
For clarity,  $c=1$ is adopted in the following discussion.

%However, the latest research shows that the important role of nonlinearity in early universe. This nonlinearity is manifested not only in the turbulence of matter \cite{caprini2009stochastic,neronov2021nanograv}, but also in the turbulence of spacetime \cite{galtier2017turbulence}. This may mean that the universe after the inflation is not as flat as expected.
%Turbulence is a form of nonlinear and non-equilibrium phenomena. In addition to motion of fluid, it is possible for any nonlinear dynamic system to exhibit turbulent behaviors (for example, turbulent black holes \cite{yang2015turbulent}). 
%Our understanding of turbulence is limited due to its complexity, but it is very common in the universe. 
%We consider the formation mechanism of primordial black holes from this natural and universal nonlinear phenomenon, which avoids going beyond the standard model or making additional physical assumptions.
%The Reynolds number for high-energy matter in the early universe is believed to be very high \cite{giovannini2012reynolds}.  
   %In this paper, we propose  turbulence as a new source  of primordial black holes, based on this natural physical processes to meet the conditions of dark matter and without presupposing any hypotheses beyond the standard model. By calculating the abundance and mass distribution of primordial black holes, we find that this result basically satisfies the constraint of primordial black holes as dark matter, and gives a strong support for primordial black holes as dark matter candidates.

\textit{MHD Turbulences and density fluctuations.--} 
For hydrodynamic turbulences, the violent density fluctuations and their implications in astrophysics had been extensively studied \cite{kaplan1970interstellar,kowal2007density}, for example the small-scale structure of electron density fluctuations \cite{armstrong1995electron} and higher order structure functions of interstellar turbulences \cite{falgarone2005intermittency}, topologies of density in interstellar mediums \cite{mckee1977theory}, and so on. 
Density fluctuations produced by compressible MHD turbulences could also play important roles in star and galaxy formation \cite{mckee2002massive,mac2004control}.
In the very early universe, indirect evidence and theoretical considerations strongly suggest the existence of turbulent processes \cite{subramanian2016origin,hindmarsh2021phase,papanikolaou2023primordial,taylor2011extragalactic,vachaspati1991magnetic,baym1996magnetic,kamionkowski1994gravitational,roper2020numerical,brandenburg1996large, pen2016shocks}. 
%Given the high Reynolds number for the primordial fluids, the electroweak phase transition will inject enough energy  through bubble collisions and lead to high Mach number fluids \cite{hindmarsh2021phase}.
%Meanwhile, bubble expansions and collisions will cause non-equilibrium processes including baryogenesis and leptogenesis, which will then lead to the generations of seed magnetic fields \cite{subramanian2016origin,papanikolaou2023primordial}.
%Given the high conductivity of the primordial plasma, MHD turbulence is then expected to form.
%Observations show the existence of large-scale coherent magnetic fields in cosmic voids \cite{taylor2011extragalactic}, and the phase transitions \cite{vachaspati1991magnetic,baym1996magnetic} together with the MHD turbulence induced by the electroweak phase transition \cite{kamionkowski1994gravitational} can be a promising mechanism to explain such large-scale magnetic fields. 

%Therefore, MHD turbulence may be an iconic feature after phase transition. 

Generally, turbulence occurs when a fluid with a large Reynolds number is perturbed. Reynolds number is defined by $Re=vL/\nu$, where $v$ is fluid velocity, $L$ the characteristic length scale, and $\nu$ the kinematic viscosity. 
For the turbulent fluid induced by the phase transition, the magnitudes of $v$ and  $L$  can be estimated according to the dynamics of the corresponding phase transition, see \cite{hindmarsh2021phase} for details of the first-order electroweak phase transition.
As the temperature of the universe decreases, the Higgs field will fall into new ground states driven by thermal fluctuations \cite{coleman1977fate} or via quantum transitions \cite{linde1981fate}. %as is suggested by the mechanism of Higgs phase transition. 
Such phase transition proceeds via bubble nucleations, that small bubbles or regions in new ground states emergent randomly, then expand, collide and merge, and finally make the whole universe enter into a new symmetry breaking phase. 
During this process, collisions among bubbles will violently stir the primordial and extremely relativistic fluid.
The energy released by the phase transition is partly injected and transformed into the MHD turbulent flow, and according to the predictions of previous studies the energy density proportion in MHD turbulences $\bar{\rho}_{tur}$ against the total energy density $\delta_{tur}=\frac{\bar{\rho}_{tur}}{\bar{\rho}}$   should be in the range $10^{-4}\sim 10^{-2}$ \cite{kosowsky2002gravitational,caprini2006gravitational,gogoberidze2007spectrum,caprini2009stochastic}, here $\bar{\rho}$ denotes the total background energy density.

It is natural to consider the turbulence length-scale as the size of phase bubbles \cite{hindmarsh2021phase}
\begin{align}
    L=2v_{w}/\beta,\label{eq:scale}
\end{align}
where $\beta$ denotes the rate parameter of the electroweak phase transition and $v_{w}$ the bubble wall speed. %Meanwhile, fluid velocity is closely related to bubble wall velocity. 
According to the relationship between $v_{w}$ and the sound speed $c_s$ in the primordial plasma, there are three possible relativistic combustion cases, subsonic deflagrations \cite{hindmarsh2019gravitational}, detonations and supersonic deflagrations (hybrids) \cite{kajantie1986bubble,espinosa2010energy,giese2020model,kurki1995supersonic}. 
In detonations, the velocity of the fluid is equal to the wall speed $v=v_{w}\sim0.87$ \cite{hindmarsh2019gravitational}\cite{caprini2009stochastic}. 
While, in supersonic deflagrations, the maximum velocity of the fluid exceeds the wall speed $c_{s}<v_{w}<v$ \cite{hindmarsh2019gravitational}, which is the case for the MHD turbulences considered in this work. 
The kinematic viscosity for the primordial plasma reads $\nu=\eta/ (\rho+p)$ \cite{caprini2009stochastic}, with $\rho,\ p$ standing for the energy density and the pressure respectively,  and $\eta$ the shear viscosity. 
%The transport coefficient can be estimated by particle interaction.  
For the high temperature primordial plasma with gauge interactions, one has the form of shear viscosity $\eta=C\frac{T^3}{g^4 \ln g^{-1}}$ \cite{arnold2000transport}
where $T$ is the temperature, $g$ is the gauge coupling and $C$ a constant.  
After the electroweak phase transition, neutrinos will decouple and cause a change in the kinematic viscosity, one then has \cite{caprini2009stochastic},
\begin{equation}
    \nu(T)\approx\left\{
	\begin{aligned}
	22T^{-1}, \quad T>100 \ \mathrm{GeV},\\
	5\times 10^{8} \mathrm{GeV}^{4}T^{-5}, \quad T<100 \ \mathrm{GeV},\\
	2\times 10^{9} \mathrm{MeV}^{4}T^{-5}, \quad T<100 \ \mathrm{MeV}.\\
	\end{aligned}
        \right
        .
\end{equation}
With the above estimation, the Reynolds number for the primordial plasma after the electroweak phase transition is about $Re=10^{13}\sim10^{16}$ \cite{giovannini2012reynolds}. To summarize here, the large Reynolds number and the extremely low viscosity will support MHD turbulence to exist for many Hubble times, therefore leading to persistent and violent density fluctuations for the production of PBH DM.

For subsonic flow, the density spectrum of MHD turbulence can be obtained by dimensional analysis. In a relatively strong magnetic field, the spectrum of energy density scales similarly to that of the pressure $\sim k^{-7/3}$ ($k$ is the wave number), when given the equation of state $p\sim \rho$ \cite{biskamp2003magnetohydrodynamic}. 
For weakly magnetized MHD turbulence, the density spectrum follows the Kolmogorov law \emph{i.e.,} $\sim k^{-5/3}$ \cite{montgomery1987density}. 
However, in supersonic flows under consideration, shock waves would pile up matter within local regions and largely enhance the magnitudes of density fluctuations at certain turbulence scales, therefore the density fluctuations will no longer be proportional to the pressure fluctuations.% namely $\delta p=c_{s}\delta \rho$. 
Without loss of generality, the fluctuations in the MHD turbulence are considered to be independent of spacetime locations, the probability distribution function (PDF) of energy or mass density will satisfy a log-normal distribution \cite{passot1998density,lemaster2008density}
\begin{eqnarray}
    P(y)\mathrm{d}y = \frac{1}{\sqrt{2\pi \sigma^{2}}}\exp\left[\frac{-(y-\mu)^{2}}{2 \sigma^{2}}\right] \mathrm{d}y,\label{eq:PDF}
\end{eqnarray}
where $y=\ln(\rho_{tur}/\bar{\rho}_{tur})$ , $\sigma^{2}$ is the variance and $|\mu|=\sigma^{2}/2$. 
Numerical simulations have confirmed the validity of this relation in three-dimensional turbulent flows \cite{lemaster2008density,kowal2007density}, and it is applicable to the primordial plasma in the early Universe.
The specific form of the PDF could be obtained by solving the detailed dynamics of the corresponding MHD turbulence model.
%Generally, such PDF of density fluctuation for MHD turbulence. 
%The dynamics of ideal MHD satisfies the following system of equations 
%\begin{eqnarray*}
   % \frac{\partial}{\partial t}\rho+\nabla\cdot(\rho\mathbf{v})=0\;,\\
  %  \frac{\partial}{\partial t}(\rho\mathbf{v})+\nabla\cdot\left(\rho\mathbf{v}\otimes\mathbf{v}-\mathbf{B}\otimes\mathbf{B}+p+B^{2}/2\right)=0\;,\\
 %   \frac{\partial}{\partial t}\mathbf{B}=\nabla\times(\mathbf{v}\times\mathbf{B})\;.
%\end{eqnarray*}
Numerical studies of MHD turbulences have been carried out, please see \cite{lemaster2008density,kowal2007density,sahraoui2020magnetohydrodynamic} for details, and these studies clearly point out that the density fluctuations strongly depend on the Mach number for both weakly and strongly magnetized turbulences. 
%As the Mach number $M_{s}$ increases, the density fluctuations become more intense, and therefore $M_s$ could serve as an indicator for the intensity of the energy injected into the turbulent system. 
For  supersonic turbulent, the relation between the density variance and Mach number $M_{s}$ has been investigated and fitted based on numerical simulations \cite{price2010density,federrath2008density,federrath2008turbulent}
\begin{eqnarray}
    \sigma_{\rho/\bar{\rho}}=b M_{s}\;,
\end{eqnarray}
where $b$ is the proportionality constant. and for the log-normal distribution in Eq. (\ref{eq:PDF}), one has
\begin{eqnarray}
    \sigma^{2}=\ln (1+b^{2} M^{2}_{s})\label{eq:sigma}.
\end{eqnarray}
According to~\cite{lemaster2008density}, the optimal linear fit gives $b^{2}=0.5$.

\textit{PBH production.--} In the radiation-dominated universe, a highly overdense region could collapse into a PBH under the action of its own gravity. The evolution of overdense regions and the formation of PBHs had been extensively studied both numerically and analytically \cite{niemeyer1998near,shibata1999black,musco2005computations,polnarev2007curvature,musco2009primordial,nakama2014identifying}.
Multiple distinct mechanisms for PBH formation have been proposed. Beyond the well-studied scenario involving the gravitational collapse of large-amplitude density perturbations upon their re-entry into the cosmological horizon after inflation, the collision of bubbles nucleated during a cosmic phase transition constitutes another significant formation channel \cite{jung2024primordial,khlopov1999first,lewicki2023primordial,baker2021primordial}. In this latter mechanism, PBHs form via the compression of matter contained within a region collapsing to a scale less than the Schwarzschild radius, induced by the coalescing bubble walls \cite{jung2024primordial}. The formation process driven by supersonic turbulence operates in an analogous fashion. Here, intense compression arising from shock waves associated with the turbulent flow can compress the material within a region below the Schwarzschild radius, thereby leading to PBH formation.

We consider a spherical overdensity region characterized by a density  $\rho=(1 + \Delta)\bar{\rho}$, where $\Delta:=\frac{\rho-\bar{\rho}}{\bar{\rho}}$ is the density contrast. The mass $M$ contained within this region of radius $R$ is given by:
\begin{align}
    M = \frac{4\pi}{3}\bar{\rho} R^3(1 + \Delta)\;.
\end{align}
 The Schwarzschild radius $R_{S}$ is 
\begin{align}
    R_s = 2GM\;,
\end{align}
Gravitational collapse into a PBH becomes possible if the radius of the overdense region is smaller than its Schwarzschild radius, $R<R_s$. This necessary condition for PBH formation translates into a requirement on the density contrast:
\begin{align}
    \Delta  > \Delta _c = \dfrac{3}{8\pi G\bar{\rho} R^2} - 1\;,
\end{align}
When considering the critical density $\bar{\rho}=\frac{3H^{2}}{8\pi G}$ as the average density

\begin{align}
    \Delta_c = \left( \dfrac{1}{HR} \right)^2-1\;.
\end{align}
When the density perturbation scale $R=\frac{1}{H}$ is comparable to the Hubble horizon, a small perturbation can lead to the formation of a   PBH. The threshold for this scenario has been discussed in many studies through analytical or numerical simulations \cite{carr1975primordial,sasaki2018primordial,nadezhin1978hydrodynamics,bicknell1979formation}, and its typical value is $\Delta_c\sim0.3$. 
Crucially, the threshold $\Delta_c$ exhibits a strong dependence on the scale of the overdensity relative to the horizon. When the overdense region is significantly smaller than the Hubble horizon $\Delta_c\sim\frac{1}{R^{2}}$.
This scaling implies that $\Delta_c$ increases sharply as the perturbation scale $R$ decreases relative to the horizon size, making PBH formation substantially more difficult for sub-horizon perturbations, in striking contrast to the near-horizon case where the threshold is less prohibitive.

%In line with the literature, given the background Friedmann-Lemaitre-Robertson-Walker (FLRW) metric, one can consider an approximately spherically symmetric overdense region, and apply the so-called separate universe approach, that the metric for the overdense region could be written down as \cite{sasaki2018primordial} 
%\begin{eqnarray}
   % \mathrm{d}s^{2}=-\mathrm{d}t^{2}+a(t)^{2}e^{2\psi(r)}\delta_{ij}\mathrm{d}x^{i}\mathrm{d}x^{j}\;,
%\end{eqnarray}
%where $\psi>0$ and $\psi \rightarrow 0$ as $r=\sqrt{\delta_{ij}x^ix^j}\rightarrow \infty$. 
%The above metric can then be transformed into the form of a locally closed universe,
%\begin{eqnarray}
   % \mathrm{d}s^{2}=-\mathrm{d}t^{2}+a(t)^{2}\left[\frac{\mathrm{d}R^{2}}{1-K(R)R^{2}}+\right. \\\notag
   % \left. R^{2}(\mathrm{d}\theta^{2}+\sin^2\theta \mathrm{d}\varphi^{2})\right]\;,
%\end{eqnarray}
%where $R=re^{2\psi(r)}$, and $K$ is given by
%\begin{eqnarray}
  %  K=-\frac{\psi'(r)}{r}\frac{2+r\psi'(r)}{e^{2\psi(r)}} .
%\end{eqnarray}
%Considering the hypersurface  with $t=constant$, the the 3-curvature for the overdense region is given by
%\begin{eqnarray}
   % R^{(3)}=\frac{K}{a^{2}}\left(1+\frac{d\ln K(R)}{3d\ln R}\right)\;.
%\end{eqnarray}
%By ignoring the spatial derivative of $K$ in the leading order gradient expansion, we can obtain the equivalent Friedmann equation \cite{sasaki2018primordial}
%\begin{eqnarray}
   % H^2+\frac{K(r)}{a^2}=\frac{8\pi G}{3}\rho,
%\end{eqnarray}
%where $H=\dot{a}/a$. This equation serves as a Hamiltonian constraint on the comoving space-like hypersurface.

We first give the estimation of the mass spectrum of PBHs from the MHD turbulence.
Like most of the PBH DM proposals in the literature \cite{carr2016primordial},  we find that the masses of the PBHs generated by the MHD turbulence will concentrate in a relatively narrow range. 
As discussed previously, the characteristic mass of PBH produced is highly related to the characteristic scale at which the fluctuations peak the most. 
The general paradigm of turbulence is large-scale energy injections accompanied by small-scale dissipation. It is then natural to consider that the major energy or mass density fluctuations are concentrated or peaked around a characteristic scale, which is considered to equal to the characteristic scale $L$ defined in Eq.(\ref{eq:scale}).
This was confirmed by previous analytical and numerical studies of the density spectrums of MHD turbulences \cite{kowal2007density,molina2012density}. 
Here, without loss of generality, we consider the case that the energy or mass density fluctuations are concentrated in the scale range $L$, \text{i.e., } $R=L$. %and unlike the cases of superhorizon fluctuations, such scale of the mass density fluctuations in MHD turbulence is generally smaller than the Hubble radius. 

According to recent studies, the bubble nucleation rate $\beta=L/2v_{\omega}$ is estimated to fall in the range $5H \sim 100 H $ \cite{peter2018gravitational}. Particularly,  in \cite{cai2018energy} it is shown that the efficiency decrease if the expansion of the background is taken into account for slow phase transitions and  $\beta$ should be $\mathcal{O}(1) H$.  
Meanwhile, recent works also point out that the formation of large bubbles is a necessary condition for PBH productions \cite{jung2024primordial}. Therefore, it is natural to consider here the characteristic scale of turbulence and the Hubble radius satisfying the relation
\begin{equation}
    LH=1\sim0.1\;. \label{eq:LH}
\end{equation}
%In fact, during the electroweak phase transition the ratio between %the characteristic scale and the Hubble radius can be estimated as %\cite{peter2018gravitational}
%\begin{equation}
    %LH=1\sim0.1. \label{eq:LH}
%\end{equation}
In the following, we work out, based on numerical calculations, the abundance of the generated PBHs and examine the conditions on the relativistic fluid across different turbulence scales. We find that the PBHs formed through this mechanism could comprise a dominant fraction of DM while the required turbulence energy density is consistent with those estimations from previous studies. 
Thus, for a preliminary exploration of this novel PBH formation channel, the relation in Eq.\~(\ref{eq:LH}) provides a physically well-motivated and self-consistent ansatz.

As PBHs considered here are supposed to be formed nearly during the same period of the MHD turbulence, one can estimate their mass range from the scaling in  Eq. (\ref{eq:LH}). The mass of a PBH is usually not greater than the total mass within the Hubble $M_H$\cite{carr2016primordial}.  When the size of an overdense region is comparable to the horizon size ($LH=1$), the mass of the PBH is
\begin{equation}
M_{PBH} = \gamma M_{\text{H}} \approx 2.03 \times 10^5 \gamma \left( \frac{t}{1 \text{ s}} \right) M_{\odot}\sim10^{25}g\;,\label{eq:HM}
\end{equation}
%resulting in a narrower mass spectrum.  
where $\gamma$ is a correction factor that can be derived analytically as $\gamma\simeq0.2$ \cite{carr1975primordial}.
When the characteristic scale $LH=0.1$ , and its $\Delta_c\sim30$, the mass of the PBH is
\begin{align}
    M_{PBH}=\epsilon\gamma M_{\text{H}}=\frac{\Delta_{c}}{(LH)^3}\gamma M_{\text{H}}\sim\frac{3}{100}\gamma M_{\text{H}}\sim10^{23}g\;,
\end{align}
where $\epsilon$ is the mass ratio of small-scale PBH to PBH on the Hubble radius scale.

 we arrive at the first important conclusion that \textit{the mass range of the PBHs generated by the MHD turbulence during and after the electroweak phase transition should be}
\begin{align}
    10^{23}\ g\le M_{\mathrm{PBH}} \le 10^{25}\ g\;,
\end{align}  
\textit{which overlap with the asteroid mass range, that the
most viable open window for the PBH DM candidate under minimal
assumptions regarding the PBH physics.} For generations of PBH with mass $\sim 10^{22} g$ with rather extreme conditions is also included in the following discussions.

Secondly, we come to the estimations of the fraction $f_{PBH}$ of the PBHs compared to the total DM component in the present universe. 
%The density contrast $\Delta:=\frac{\rho-\bar{\rho}}{\bar{\rho}}$ on the comoving space-like hypersurface satisfies \cite{sasaki2018primordial},
%\begin{eqnarray}
  %  \Delta=\frac{3K}{8\pi G\bar{\rho}a^2}=\frac{K}{H^2a^2} .
%\end{eqnarray}
%There have been many discussions on the threshold $\Delta_c$ of the density contrast for PBHs productions. 
%At first glance, in order to collapse against the pressure, the energy or mass density fluctuations need to be larger than the Jeans length, which could take place when $\Delta\ge\Delta_{c}\sim c_{s}^{2}\sim\frac{1}{3}$ \cite{carr1975primordial,sasaki2018primordial}. 
%Studies based on numerical simulations of evolutions of overdense spherically symmetric regions had confirmed this conclusion \cite{nadezhin1978hydrodynamics,bicknell1979formation}. 
%More detailed research further pointed out that the threshold could also depend on the forms of the curvature profile, and $\Delta_{c}$ should fall in the range of $0.4\sim2/3$ \cite{musco2019threshold}.
The abundance parameter of the produced PBHs during the MHD turbulence period against the total energy or mass density in the universe can be defined as
 \begin{eqnarray}
 \alpha:=\frac{\rho_{\mathrm{PBH}}}{\bar{\rho}}\bigg|_{\mathrm{at~formation}}\;\nonumber,
 \end{eqnarray}
which relates to the faction parameter $f_{PBH}$ as \cite{carr2010new}
\begin{equation}
    \alpha \simeq 3.7\times 10^{-9}(\frac{\epsilon\gamma}{0.2})^{-\frac{1}{2}}(\frac{g_{*,form}}{10.75})^{\frac{1}{4}}(\frac{M_{PBH}}{M_{\odot}})^{\frac{1}{2}}f_{PBH}.\label{eq:fpbh}
\end{equation}
Here $g_{*,form}\sim 100$ stands for the total relativistic degrees of freedom at that PBH formation epoch \cite{carr2010new}. 
Here, we consider the case that the mass distribution function of PBH is monochromatic, that it is a peak-shape function that is centered at a certain $M_{PBH}$ with width $\Delta M\sim M_{PBH}$. 
Given the PDF of the density fluctuation in Eq. (\ref{eq:PDF}), $\alpha$ can be calculated according to the Press–Schechter formalism \cite{press1974formation} and it is then interpreted as the probability that the energy or mass density fluctuation is larger than the given threshold
 \begin{equation}\alpha=\gamma\int_{y_{\mathrm{th}}}^\infty P(y) \mathrm{d}y.
 \end{equation}

Here the integration lower bound is given by $y_{th}=ln(1+\frac{\Delta_{c}}{\delta_{tur}})$.
This is because, as mentioned, only a small fraction of total energy is converted into turbulent fluctuations $\bar{\rho}_{tur}=\bar{\rho}\delta_{tur}$, implying that these overdense regions are sparsely distributed. 
The turbulence fluctuations exceeding the threshold $y>ln(\frac{\bar{\rho}\delta_{tur}+\bar{\rho}\Delta_{c}}{\bar{\rho}\delta_{tur}})=ln(1+\frac{\Delta_{c}}{\delta_{tur}})$ with respect to the energy density $\rho_{tur}$  
is counted in the above equation. 
%is obtained from the PBH formation threshold $\Delta_c$ and the energy density proportion in turbulences $\delta_{tur}$. 
With the log-normal distribution of MHD turbulence in Eq. (\ref{eq:PDF}) - (\ref{eq:sigma})  and the relation in Eq. (\ref{eq:fpbh}), we can then access how the fraction $f_{PBH}$ of PBHs in the present universe depends on the related parameters of PBH formation and the MHD turbulence, including the peak PBH mass $M_{PBH}$, the formation threshold $\Delta_c$, the proportion of energy density in turbulence $\delta_{tur}$ and the Mach number $M_s$. 
%The relationship between variance and Mach number $M_{s}$ is $\sigma^{2}=\ln (1+0.5 M^{2}_{s})$. 

%\begin{figure*}
%\includegraphics[width=0.48\linewidth]{0.31.png}
%\includegraphics[width=0.48\linewidth]{0.41.png}
%\includegraphics[width=0.48\linewidth]{0.51.png}
%\includegraphics[width=0.48\linewidth]{0.61.png}
%\caption{\label{fig} The four figures show  the estimations of the abundance of PBHs formed during the MHD turbulence period induced by the electroweak phase transition, given different threshold $\Delta_{c}=0.3, 0.4, 0.5, 0.6$. For the extremely relativistic and supersonic flow, the sound speed is $1/\sqrt{3}$ and Mach number $M_{s}$ takes value from 1 to 1.5. The curves on the graph are labeled with the abundances $\alpha=10^{-13},10^{-14},10^{-15},10^{-16}$ and   the regions between the curves correspond to the fraction of dark matter contributed by PBHs being $f_\mathrm{PBH}\sim1,\frac{1}{10},\frac{1}{100}$ .} 
%\end{figure*}

\begin{widetext}
\begin{figure*}[ht]
  \subfloat[]{\includegraphics[width=0.27\textwidth]{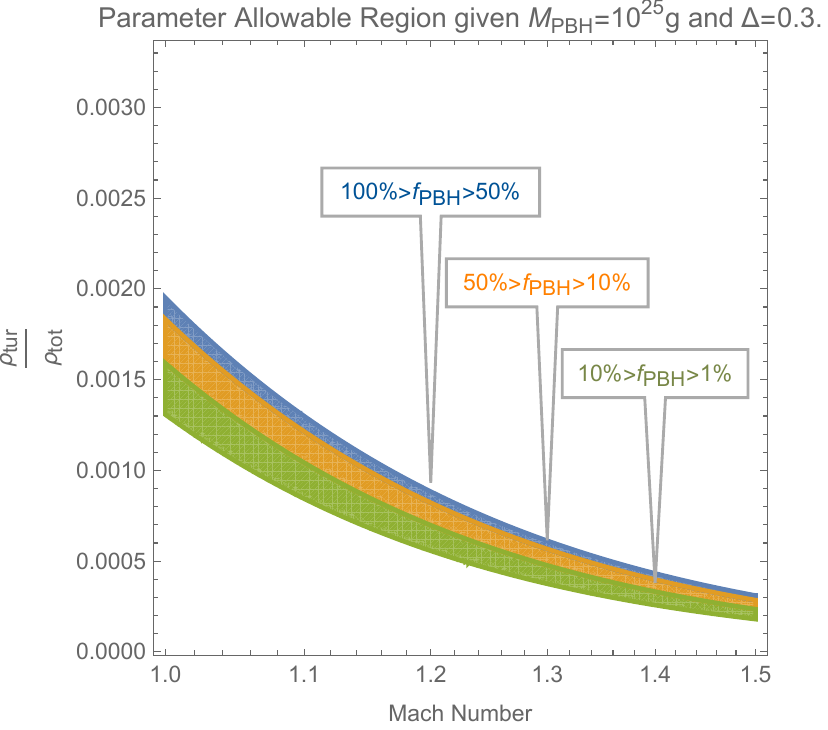}\label{fig:Ma}}
  \subfloat[]{\includegraphics[width=0.255\textwidth]{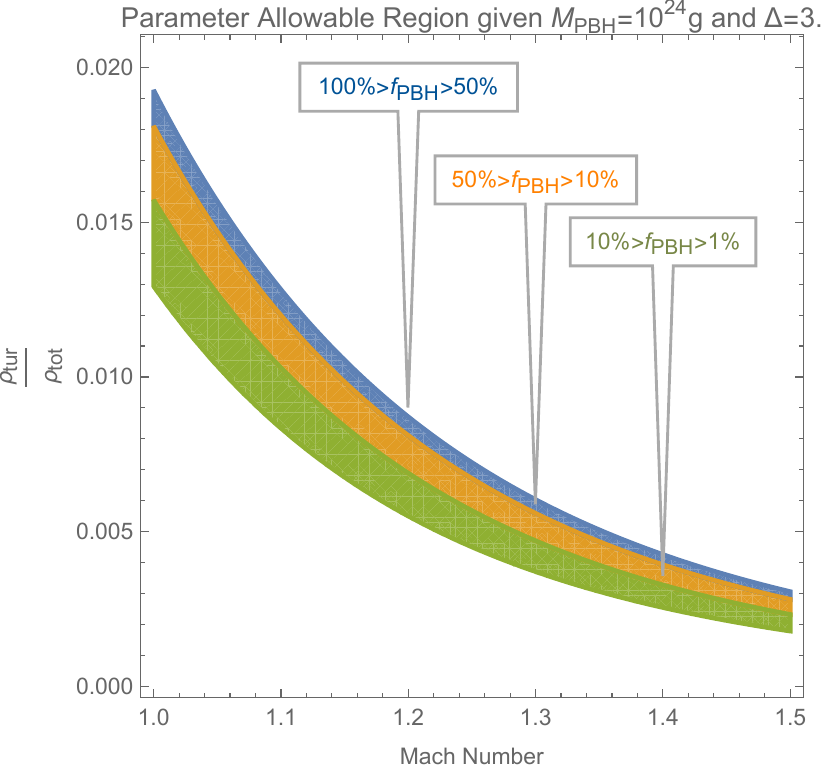}\label{fig:Mb}}
  \subfloat[]{\includegraphics[width=0.25\textwidth]{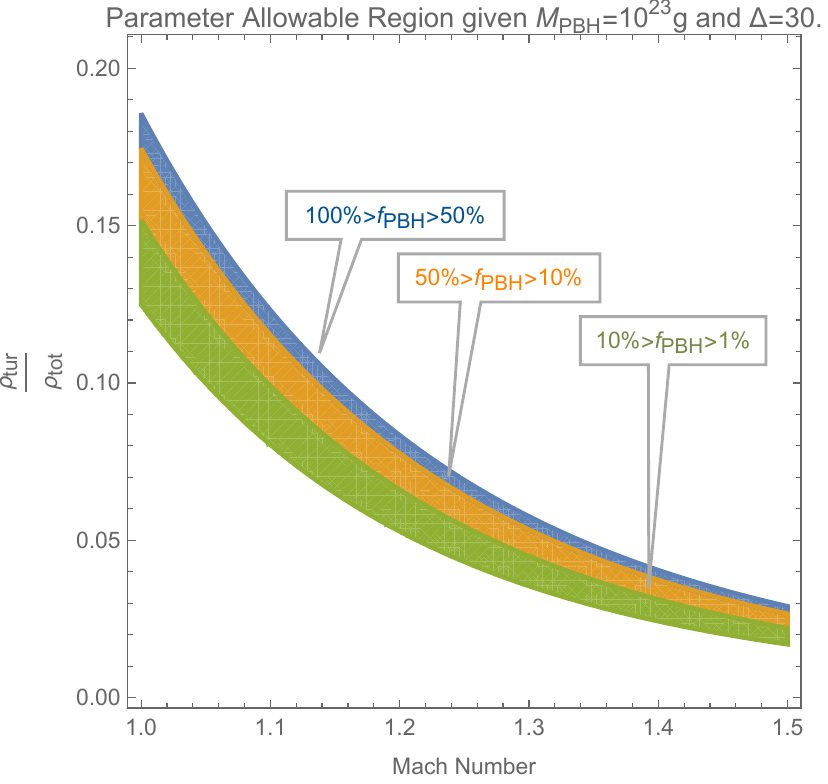}\label{fig:Mc}}
   \subfloat[]{\includegraphics[width=0.245\textwidth]{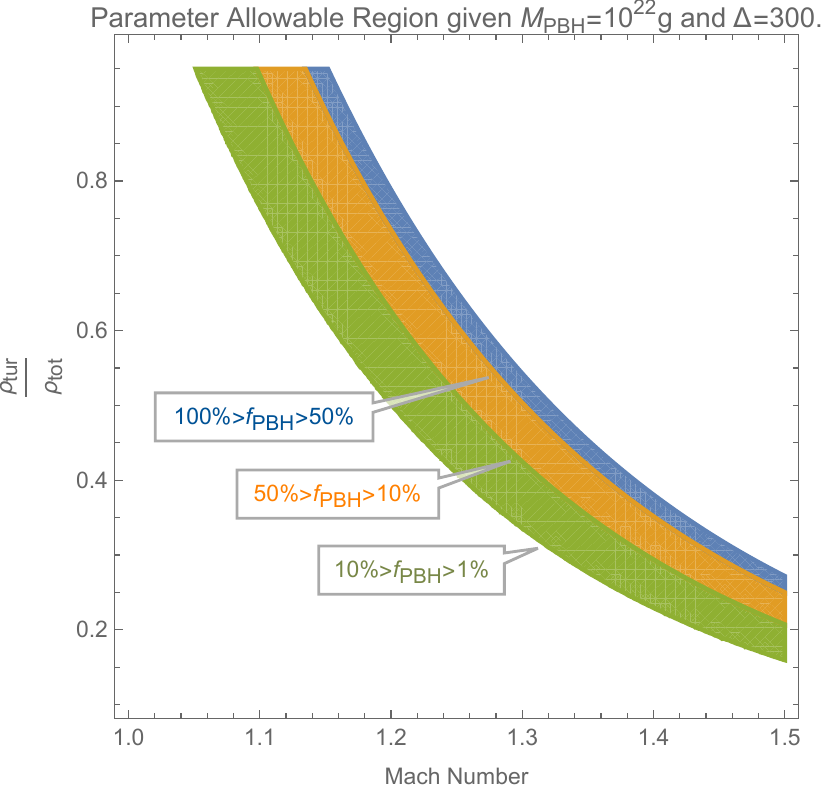}\label{fig:Mc}}
  \caption{We consider four representative cases of the turbulence characteristic scale $LH=1,\ \sqrt{\frac{1}{10}},\ \frac{1}{10}, \frac{1}{10\sqrt{10}}$,  which correspond to the thresholds 
$\Delta_c=$0.3, 3, 30, 300 and the PBH masses $10^{25}g,10^{24}g,10^{23}g,10^{22}g $. The allowable regions in the $M_{PBH}-\delta_{tur}$ (peak PBH mass-energy density proportion in turbulences) parameter space for PBH formation with the abundance $f_{PBH}$ within $50\%\sim100\%$, $10\%\sim 50\%$ and $1\%\sim 10\%$ are shown.}
  \label{fig:M-rho}
\end{figure*}

%\begin{figure*}[ht]
  %\subfloat[]{\includegraphics[width=0.32\textwidth]{Ms-rho-3.pdf}\label{fig:Msa}}
 % \subfloat[]{\includegraphics[width=0.32\textwidth]{Ms-rho-5.pdf}\label{fig:Msb}}
 % \subfloat[]{\includegraphics[width=0.32\textwidth]{Ms-rho-6.pdf}\label{fig:Msc}}
  %\caption{Based on the discussions in this work, we set the peak mass of the formed PBH to a typical value of $M_{PBH}=10^{22}\ g$, and consider three representative cases of the threshold $\Delta_c=$0.3, 0.5, and 0.6. The allowable regions in the $M_{s}-\delta_{tur}$ (Mach number - energy density proportion in turbulences) parameter space for PBH formations with the abundance $f_{PBH}$ within $50\%\sim100\%$, $10\%\sim 50\%$ and $1\%\sim 10\%$ are shown.}
  %\label{fig:Ms-rho}
%\end{figure*}

As shown in Fig. \ref{fig:M-rho}, for the three representative cases with $LH=1,\ \sqrt{\frac{1}{10}},\ \frac{1}{10}, \frac{1}{10\sqrt{10}}$, which correspond to the thresholds 
$\Delta_c=$0.3, 3, 30, 300 and the PBH masses $10^{25}g,\ 10^{24}g,\ 10^{23}g, 10^{22}g$, we give the allowable regions in the  $M_{s}-\delta_{tur}$ parameter spaces respectively for PBH fraction $f_{PBH}=50\%\sim100\%$, $10\%\sim 50\%$ and $1\%\sim 10\%$.  
The value $\Delta_{c}=0.3$ corresponds to a larger characteristic turbulent scale and is associated with the formation of large-scale bubbles. 
This scenario is dynamically well-motivated with a relatively high probability in the context of strong first-order phase transitions, similar to the conclusion in \cite{jung2024primordial} that large bubble collisions are essential for efficient PBH production. As such, $\Delta_{c}=0.3$ represents the primary physical scenario considered in this work.
On the other hand, the cases $\Delta_{c}=3$ and $\Delta_{c}=30$ correspond to the effective density thresholds defined such that the overdense region collapses within a physical scale smaller than the Hubble radius, where such smaller-scale structures are  potentially generated from shock compression in highly supersonic turbulent flows. 
Calculations indicate that the high threshold scenarios require prohibitively intense turbulent flows, thereby exhibiting low possibility.
These high-threshold cases are included to ensure the completeness of our analysis which may constitute subdominant channels.
Here, the Mach number of the supersonic MHD turbulence is considered to take a value near 1, since for the extremely or even fully relativistic primordial MHD turbulence the Mach number could not be much larger than 1 \cite{gogoberidze2007spectrum}. Meanwhile, simulations of relativistic combustion show that the phase transition bubble-wall speed is between $\frac{1}{\sqrt{3}}c$ and $0.87c$ \cite{hindmarsh2021phase}.
Therefore, in Fig. \ref{fig:M-rho}, the Mach number is set to a typical value of $1-1.5$, and within the asteroid mass range one finds that to comprise a dominant part of DM the proportion of energy density in turbulences should fall into the range of $10^{-4}\sim 10^{-2}$, which is remarkably in high agreement with the previous and independent studies on how much energy was injected and transformed into MHD turbulent flow during and after the electroweak phase transition \cite{kosowsky2002gravitational,caprini2006gravitational,gogoberidze2007spectrum,caprini2009stochastic}. Fig. \ref{fig:M-rho} clearly shows that as the scale of overdense regions diminishes, a greater proportion of turbulent energy and a higher Mach number are needed to form PBH.

%This holds true for other reasonable choices of the Mach number, as demonstrated in Fig. \ref{fig:Ms-rho}.
%In Fig. \ref{fig:Ms-rho}, we set the typical value of the peak PBH mass as $10^{22}\ g$, and, as expected, the abundance of PBHs  is not sensitively dependent on the Mach number of the turbulent plasma. 
%These results in fact impose an independent constraint upon the energy density proportion of the primordial MHD turbulence, $\delta_{tur}=10^{-4}\sim10^{-3}$, in forming a dominant PBH DM. 

\end{widetext}

Here, we obtain the second important and rather robust conclusion that, \textit{without the employments of fine-tuned dynamics of primordial quantum fields and inflation models, the supersonic MHD turbulence taken place during and after the electroweak 1st-order phase transition in the very early universe can generate PBHs comprising a dominant part of DM, and the constraint upon how much energy was transformed into turbulences matches remarkably with the previous and independent studies.}

%The mass range of sublunar black holes, spanning from $10^{20}$ grams to $10^{24}$ grams \cite{carr2016primordial}, aligns closely with our findings. 
%Considering pertinent constraints stemming from microlensing observations, dynamical impacts, large-scale structural analyses, and gravitational waves detected by the LIGO-Virgo collaboration \cite{carr2016primordial}, it is plausible that sublunar black holes may account for the entirety of dark matter.  

Therefore, the above analysis of the abundance and the mass range of the PBHs gives us a very strong indication that the important influences and implications of turbulences in the very early universe should not be overlooked and the MHD turbulence induced by the electroweak phase transition could provide a natural mechanism to generate the right DM candidate that satisfies the present-day observational constraints. Additionally, in this work we adopt the well-accepted electroweak phase transition energy scale of $100\ GeV$, corresponding to a cosmic time of \(\sim 10^{-12}\) s. 
If the transition occurred slightly earlier, the horizon mass would decrease, resulting in the formation of lower-mass PBHs. We therefore consider the possibility of forming lighter PBHs, down to \(10^{22}\) g, which could arise on smaller turbulent scales (see Fig.~\ref{fig:M-rho}-(d)). While this scenario requires significantly enhanced turbulent energy density and represents an extreme case, it remains physically plausible under sufficiently intense MHD turbulence induced by a slightly earlier phase transition. By including this possibility, we demonstrate the flexibility of the formation mechanism across a broad mass spectrum, which may bring the predicted abundance into better agreement with observational constraints \cite{carr2021constraints,niikura2019microlensing}.

\textit{Concluding remarks.--}PBHs have become an important subject in modern astronomy, cosmology, and also fundamental physics, offering a compelling candidate for elucidating the origin and nature of DM. The production of PBHs through well-established and accepted physical mechanisms is highly worthy of in-depth investigations.  This work highlights the role of turbulences in the very early universe, as a natural state of extremely relativistic fluids subjected to violent disturbances from phase transitions, in sustaining intense and persistent fluctuations in energy or mass density. Such turbulent behavior provides a natural mechanism for PBH formation in the primordial universe. Based on previous studies of the electroweak phase transition, we analyze the abundance and mass range of PBHs produced by the MHD turbulence it induces. Remarkably, we find that the mass range of the PBHs falls precisely within the most viable asteroid mass window provided by current joint observations, and to form a dominant fraction of asteroid mass PBH DM an independent but natural constraint  $\delta_{tur}=10^{-4}\sim10^{-2}$ on the energy density proportion of the MHD turbulence is derived. This new constraint on the energy proportion of MHD turbulence closely matches that from previous studies. Such findings suggest that turbulence in the early universe, particularly MHD turbulence induced by the electroweak phase transition, could play a critical role in the generation of PBHs as a dominant part of DM.  
%Upon further consideration, other sources of turbulence caused
%turbulence in the early universe, such as reheating \cite{subramanian2016origin} and QCD phase transition, may also give rise to PBHs. In addition, we believe that turbulence will intensify the accretion process of PBHs, which may also lead to their greater abundance. This may further increase the abundance and mass range of PBHs.

However, our analysis is primarily based on qualitative analysis and order-of-magnitude estimates, future studies will focus on refining the formation mechanisms and dynamical evolution of PBHs to further solidify this compelling scenario. Turbulence in the primordial plasma may also arise from different nonlinear processes in the early Universe, such as those during reheating \cite{micha2003relativistic,micha2004turbulent} and the QCD phase transition \cite{khachatryan2008modified}. Reheating can drive turbulent thermalization of classical wave fields, initiating relativistic hydrodynamic cascades. During the first-order QCD phase transition, bubble collisions and plasma shocks may generate highly supersonic flows, resulting in high Mach number turbulence. Conversely, PBHs themselves may act as seeds for primordial magnetic fields through mechanisms such as accretion-driven plasma instabilities or anisotropic emission processes \cite{papanikolaou2023primordial,safarzadeh2018primordial}. Together, these seemingly interrelated processes highlight the early Universe as a rich environment for nonlinear hydrodynamical and magnetohydrodynamical phenomena, that closely tied to PBH formation and evolution.

\begin{acknowledgments}
This work is supported by the International Partnership Program of the Chinese Academy of Sciences, Grant No. 025GJHZ2023106GC.
\end{acknowledgments}

\appendix

\nocite{*}

\bibliography{apssamp}% Produces the bibliography via BibTeX.

\end{document}